\documentclass[review]{elsarticle}
\usepackage{multirow,tabularx}

\usepackage{journalnames}
\usepackage{hyperref}

\journal{Icarus}

\biboptions{authoryear}

\begin{document}

\begin{frontmatter}

\title{Assessing the long-term variability of acetylene and ethane in the stratosphere of Jupiter}

\author[leicester]{Henrik Melin}
\author[leicester]{L. N. Fletcher}
\author[leicester]{P. T. Donnelly}
\author[swri]{T. Greathouse}
\author[austin]{J. Lacy}
\author[jpl]{G. S. Orton}
\author[jpl]{R. Giles}
\author[jpl]{J. Sinclair}
\author[oxford]{P. G. J. Irwin}

\address[leicester]{University of Leicester, Leicester LE1 7RH, UK}
\address[swri]{Southwest Research Institute, San Antonio, USA}
\address[austin]{University of Texas, Austin USA}
\address[jpl]{Jet Propulsion Laboratory, Pasadena, USA}
\address[oxford]{University of Oxford, UK}

\begin{abstract}
Acetylene (C$_2$H$_2$) and ethane (C$_2$H$_6$) are both produced in the stratosphere of Jupiter via photolysis of methane (CH$_4$). Despite this common source, the latitudinal distribution of the two species is radically different, with acetylene decreasing in abundance towards the pole, and ethane increasing towards the pole. We present six years of NASA IRTF TEXES mid-infrared observations of the zonally-averaged emission of methane, acetylene and ethane. We confirm that the latitudinal distributions of ethane and acetylene are decoupled, and that this is a persistent feature over multiple years. The acetylene distribution falls off towards the pole, peaking at $\sim$30$^{\circ}$N with a volume mixing ratio (VMR) of $\sim$0.8 parts per million (ppm) at 1 mbar and still falling off at $\pm70^\circ$ with a VMR of $\sim$0.3 ppm. The acetylene distributions are asymmetric on average, but as we move from 2013 to 2017, the zonally-averaged abundance becomes more symmetric about the equator. We suggest that both the short term changes in acetylene and its latitudinal asymmetry is driven by changes to the vertical stratospheric mixing, potentially related to propagating wave phenomena. Unlike acetylene, ethane has a symmetric distribution about the equator that increases toward the pole, with a peak mole fraction of $\sim$18 ppm at about $\pm50^{\circ}$ latitude, with a minimum at the equator of $\sim$10 ppm at 1 mbar. The ethane distribution does not appear to respond to mid-latitude stratospheric mixing in the same way as acetylene, potentially as a result of the vertical gradient of ethane being much shallower than that of acetylene. The equator-to-pole distributions of acetylene and ethane are consistent with acetylene having a shorter lifetime than ethane that is not sensitive to longer advective timescales, but is augmented by short-term dynamics, such as vertical mixing. Conversely, the long lifetime of ethane allows it to be transported to higher latitudes faster than it can be chemically depleted.


\end{abstract}

\begin{keyword} \\
{\bf
Jupiter, atmosphere \\
Spectroscopy \\
Abundances, atmospheres \\
Atmospheres, structure \\
}
\end{keyword}

\end{frontmatter}

\section{Introduction}

The temperature of Jupiter's stratosphere is determined by the balance between solar heating via absorption of sunlight in the visible and near-infrared, and radiative cooling by hydrocarbons, predominantly in the mid-infrared \citep{2001Icar..152..331Y, 2004jpsm.book..129M}. 

Photolysis of methane (CH$_4$) produces acetylene (C$_2$H$_2$) and ethane (C$_2$H$_6$). And all three of these species are efficient emitters in the mid-infrared and cool the volume of gas that they inhabit by radiating energy to space \citep{2004jpsm.book..129M}. By observing the spectra of these species, we can retrieve the vertical temperature profiles, and the acetylene and ethane abundances \citep{2008JQSRT.109.1136I}. This gives us our primary means of diagnosing environmental conditions in Jupiter's stratosphere, revealing the dynamical and chemical processes at work.



Acetylene and ethane were first discovered at Jupiter by \cite{1974ApJ...187L..41R} using the 1.5 m McMath solar telescope at Kitt Peak. The first latitudinal profiles were derived from Voyager-2 IRIS spectra by \cite{1984BAAS...16..647M}, showing that the zonal abundances of acetylene and ethane were radically different: the mixing ratio of the former decreased by a factor of three towards the pole, whilst the latter increased by a factor of three towards the pole. This was confirmed by \cite{2007Icar..188...47N}, who analysed mid-infrared observations taken with the CIRS instrument \citep{2004SSRv..115..169F} during the Cassini flyby of Jupiter in 2000. These two abundance distributions are surprising, given that both species are photochemical products of methane, and  have an expected peak production rate about the equator, where the solar flux is the greatest \citep{2004jpsm.book..129M}. 

Very few studies constrain the long-term change in the abundance of ethane. \cite{1987Icar...72..394K, 1989InfPh..29..199K} analyzed data from 1982 to 1986 and observed no change in the ethane abundance. More recently, \cite{2010P&SS...58.1667N} analysed mid-infrared acetylene and ethane emissions from both the Voyager and Cassini spacecraft, providing two snapshots of their distribution separated by 21 years, sampling northern fall equinox and northern summer solstice respectively. They observed an ethane abundance that increased slightly and became more symmetric between 1979 and 2000. In contrast, the acetylene abundance went from being almost constant with latitude to having an abundance distribution that fell sharply towards the pole. They concluded that there could be seasonal changes in temperature, acetylene, and ethane at all latitudes and pressure levels. Improvements in ground-based spectroscopic mapping capabilities were exploited by \cite{2016Icar..278..128F} to provide a further snapshot of Jupiter's stratospheric temperatures, ethane and acetylene in 2014, finding very similar distributions to those seen by Cassini 14 years earlier.  In this work, we expand on the work of \cite{2016Icar..278..128F} to study stratospheric variability over a 6-year timescale, or half a jovian year.

\cite{2010P&SS...58.1667N} modelled the lifetimes at 1 mbar to be hundreds of days for acetylene and hundreds of years for ethane. The former is much shorter than the jovian year ($\sim$12 years) whilst the latter is likely too long to sense any short-term seasonal or dynamic behaviour. These two very different timescales could explain why the distribution of ethane and acetylene appear so different. If dynamic timescales are either much shorter or much longer than these lifetimes we might expect the distributions to be identical. If, however, the typical dynamical timescales are in-between the two lifetimes then we may expect the two species to show different zonally averaged meridional distributions \citep{2013ApJ...767..172Z, 2013Icar..225..228M, 2014Icar..242..149K}. However, \cite{2005ApJ...635L.177L} suggested that these arguments based on chemical lifetimes alone were likely too simplistic. They used a coupled photochemical and dynamical model of Jupiter's stratosphere to show that whilst different acetylene and ethane lifetimes could produce different zonal distributions, they required stratospheric transport timescales at 5 mbar that were 100 times longer than those derived from the Shoemaker-Levy 9 impact observations \citep{1999Icar..138..141F}. 


At the poles of Jupiter, methane chemistry is also driven by energetic auroral precipitation \citep[e.g.][]{2017Icar..292..182S}. This generates enhancements in ethane and acetylene about the magnetic poles, which rotate at the System III longitude rate. These components may contribute to the global distribution, but the mechanisms for this remain unclear, and the present study will focus on the equatorial and mid-latitudes away from the auroral regions.

One persistent long-term dynamic feature of Jupiter's stratosphere is the Quasi-Quadrennial Oscillation \citep[QQO, ][]{1991Natur.354..380L, 1991Sci...252..537O, 1994Sci...265..625O}, whereby the temperature of the equatorial stratosphere oscillates with an approximate four-year cycle. This is observed as changes in the zonal mean brightness temperature of methane at any particular pressure level. An increase in stratospheric temperatures will increase the observed brightness of acetylene and ethane, but it is unknown if the QQO has a direct role to play in driving long-term changes in stratospheric abundances. 


Here, we investigate how stratospheric temperatures and the distributions of acetylene and ethane change over a period of six years, between September 2012 and January 2017. We describe our observations in Section \ref{secobs}, followed by the testing of a variety of different plausible spatial distributions using atmospheric retrievals in Section \ref{secanalysis}. The results are described in Section \ref{secresults}, followed by a discussion, placing them in a broader context in Section \ref{secdiscussion}. We then summarise the results of this study in Section \ref{secsummary}.

\begin{table}[]
\centering

\begin{tabular}{l | ll | ll}
TEXES Run & \multicolumn{2}{c|}{Included?}  &  & Comment  \\
\hline
\hline
2012 January 		&    		& No &  &   Poor weather \\
2012 September 	& Yes  	&   &  &  Only methane band \\
2013 February 		& Yes  	&  &  &  \\
2014 February 		&   		& No  &  &  Poor weather \\
2014 December 	& Yes  	&  &  &   \\
2015 March 		& Yes  	&   &  &  \\
2015 November	& Yes  	&  &  &  \\
2016 January 		&  		& No  &  & Instrument issues  \\
2016 April & Yes  	&  &  &  \\
2016 December 	& Yes  &  &  &   \\
2017 January 		& Yes   &  &  & \\
\hline
\end{tabular}
\caption{The NASA IRTF TEXES observing runs used in this study. Note that the 2014 December observations were presented by \cite{2016Icar..278..128F}. This usability implied by this table only applies to the medium resolution observations analysed in this study. \label{observations}}
\end{table}

\begin{table}
\centering
\begin{tabular}{ llll }
\hline
Observing run & 744 cm$^{-1}$ scaling & 819 cm$^{-1}$ scaling & 1247 cm$^{-1}$  \\
\hline
\hline

2012 September & N/A & N/A & 1.30 \\
2013 February & 1.10 & 0.73 & 1.00 \\
2014 December & 1.06 & 0.73 & 1.00 \\
2015 March & 1.09 & 0.73 & 1.00 \\
2015 November & 1.04 & 0.68 & 1.00 \\
2016 January & 1.12 & 0.74 & 1.00 \\
2016 April & 1.12 & 0.74 & 1.00 \\
2016 December & 1.09 & 0.77 & 1.00 \\
2017 January & 1.48 & 0.93 & 1.30 \\

 \hline
\end{tabular}
\caption{The multiplicative scaling factor for the acetylene, ethane, and methane bands for each observing run used in this study. This is derived by comparing the mid-latitude tropospheric continuum emissions of NASA IRTF TEXES and Cassini CIRS. This scaling factor was applied to the TEXES observations and are plotted in Figure \ref{scalings}.}
 \label{tscalings}
\end{table}

\section{Observations \label{secobs}}


The Texas Echelon Cross Echelle Spectrograph \cite[TEXES,][]{2002PASP..114..153L} is a mid-infrared spectrograph capable of observing at spectral resolutions between $R= \lambda / \Delta \lambda \sim2000$ to $R\sim85$,000, used in a visitor mode at the 3-m NASA Infrared Telescope Facility (IRTF) in Hawaii. The instrument covers the wavelength range 5 -- 25 $\mu$m at a variety of slit-lengths and spectral resolutions -- here we use the 45$^{\prime\prime}$-long slit (1.4$^{\prime\prime}$ wide in medium-resolution mode), giving a spectral resolution up to  $R \sim$10,000. The long slit covers the entire diameter of Jupiter, which has an maximum angular size at opposition of $\sim$45$^{\prime\prime}$. The TEXES detector contains 256 $\times$ 256 pixels. 

We use three medium-resolution settings: acetylene at 744 cm$^{-1}$ (13.4 $\mu$m, coverage between 742-747 cm${-1}$), ethane at 819 cm$^{-1}$ (12.2 $\mu$m, coverage between 815-823 cm$^{-1}$), and methane at 1247 cm$^{-1}$ (8.0 $\mu$m, coverage between 1243-1252 cm$^{-1}$). The methane emissions sense stratospheric temperature, whilst the acetylene and ethane bands enable the determination of stratospheric hydrocarbon abundances. From 2013 onwards the TEXES observations  fell into a regular pattern of acquiring eight spectral channels in medium and low resolution:  538 and 587 cm$^{-1}$ in the Q-band, 745, 819 and 1248 cm$^{-1}$ for stratospheric emission; and 900, 960 and 1160 cm$^{-1}$ for tropospheric structure. The motivation for choosing these channels is explained in more detail by \cite{2016Icar..278..128F}. 

The slit was aligned along the north-south terrestrial meridian, and was stepped across the disc of Jupiter, building up a spectral cube of two spatial dimensions and one spectral dimension. At the beginning and end of each spectral cube, calibration flat-field and black-body spectra were obtained by moving a card of known temperature in front of the instrument aperture. Typical exposure times are $\sim$4 s per spectrum, with a full scan of Jupiter taking about 5 minutes. 

These observations are wavelength calibrated using the telluric skylines and flux calibrated using the calibration black body observations and the TEXES `{\tt pipe}' pipeline software \citep{2002PASP..114..153L}. Once calibrated, the calculated projection of the limb of Jupiter was fitted to the observations by eye, enabling the allocation of planetographic latitude, System III longitude, and emission angle  to each pixel. This information can be used to project the calibrated spectral cube onto a latitude and longitude grid, which in turn enables us to average several spectral cubes together to build up complete latitude--longitude maps for each wavelength. 



\begin{figure}
\centering
\includegraphics[width=0.8\textwidth]{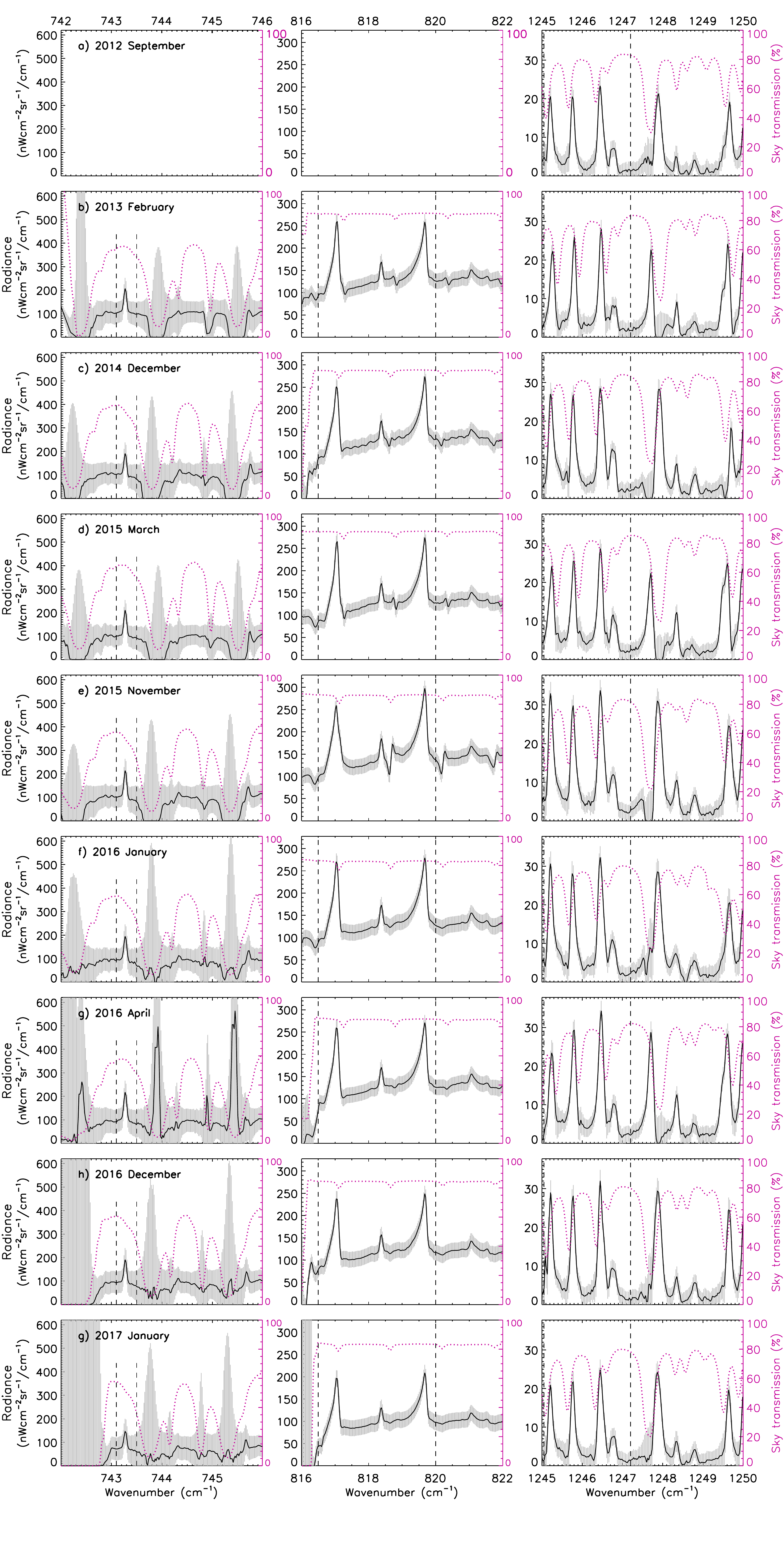}
\caption{ The equatorial spectra of Jupiter for the epochs listed in Table \ref{observations}, showing the three wavelength bands: 744 cm$^{-1}$, 819 cm$^{-1}$, and 1247 cm$^{-1}$ (left to right). The black line is the observed intensity, the grey shaded regions is the uncertainty on the intensity and the purple line is the telluric absorption. The dashed lines shows the wavenumber range used for the NEMESIS retrievals. The spectra have been shifted to rest wavelengths.\label{all_spectra}}
\end{figure}

\begin{figure}
\centering
\includegraphics[width=1.0\textwidth]{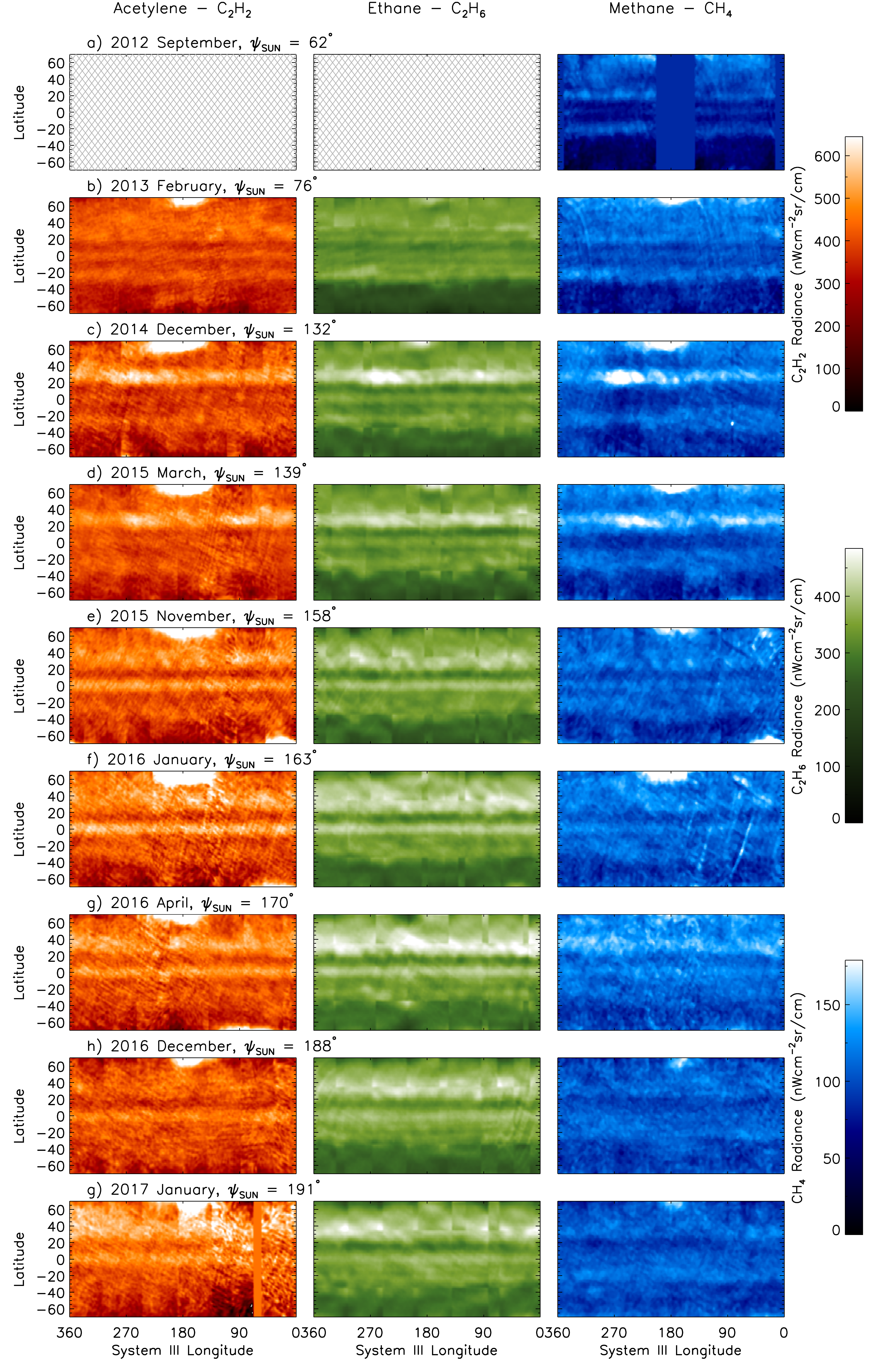}
\caption{From left to right, maps of acetylene, ethane, and methane emissions extracted from the 744, 819 and 1247 cm$^{-1}$ medium spectral resolution channels on the TEXES instrument, respectively. The latitudes are planetographic and the longitudes are System III. All the channels have been scaled as to show the observed intensity at all latitudes. The solar ecliptic longitude $\psi_{SUN}$ is indicated, with $\Psi_{SUN}$ = 90$^\circ$ representing northern summer solstice. \label{maps}}
\end{figure}

We analyse the acetylene, ethane, and methane TEXES observations of Jupiter obtained between 2012 and 2017, listed in Table \ref{observations}. Because of weather and technical issues, not all observing runs produced data usable for this study. The September 2012 observations had very limited longitude coverage for the acetylene and ethane bands, but we were able to construct a near-complete map of methane, allowing us to retrieve temperature. The January 2016 observations suffered from a partial obstruction of the light path of the instrument by mylar, following a routine servicing of the instrument. The  equatorial TEXES spectra from the observing runs, averaged between 30$^{\circ}$N and 30$^{\circ}$S, are shown in Figure \ref{all_spectra}. The longitude coverage is complete for all runs, apart from September 2012 (see Figure \ref{maps}).


\begin{figure}
\centering
\includegraphics[width=0.9\textwidth]{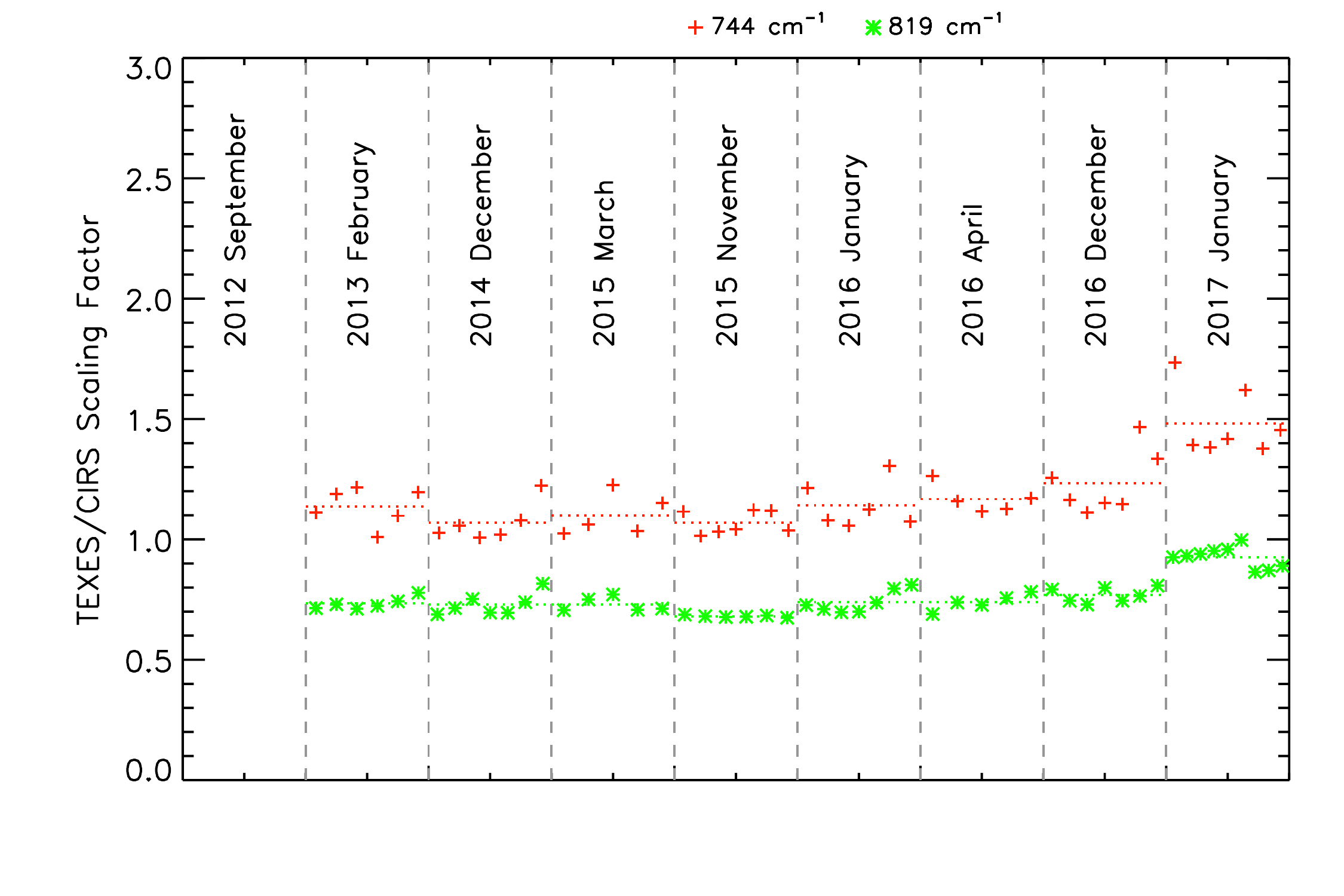}
\caption{ The intensity scaling factor used for the TEXES observations analysed here. Each cross represents an individual acetylene data cube, whilst stars indicate individual ethane cubes. The factor is approximately the same for all years, apart from January 2017. These factors were derived by comparing the intensity of the tropospheric mid-latitude emission to that of Cassini CIRS. \label{scalings}}
\end{figure}


The TEXES detector displays non-linear behaviour that is a function of the energy flux that falls on the detector. Since the energy flux is dependent on the spectral resolution, this becomes a noticeable issue at medium and low resolutions, and in spectral windows where the target has a large energy flux. At Jupiter, the 744 cm$^{-1}$ window is brighter than the 1247 cm$^{-1}$ window by about a factor of 10 per spectral pixel, which is why we need to introduce a larger correction factor at 744 than at 1247 cm$^{-1}$. \cite{2016Icar..278..128F} addressed this issue in the December 2014 data by introducing a scaling factor, and a similar approach will be adopted here. In order to correct for this detector behaviour we scale the TEXES spectra using the continuum emissions from the troposphere, matching it to a TEXES forward model based on CIRS temperatures and compositions. The Cassini CIRS instrument is one of the most accurately calibrated mid-IR spectrometers \citep{Jennings:17}, and a comparison between the two instruments can be illustrative of inconsistencies in the TEXES calibration. 


To determine the radiance of the tropospheric background, we select a region void of any stratospheric emissions and telluric absorption features. For the 744 cm$^{-1}$ band we use the region between 743.0 and 743.1 cm$^{-1}$ (left of the acetylene feature in Figure \ref{all_spectra}). For the 819 cm$^{-1}$ band we use the region between 817.4 and 818.1 cm$^{-1}$ (between the ethane features in Figure \ref{all_spectra}). We compare the latitude regions contained within 45$^{\circ}$N to 60$^{\circ}$N and 45$^{\circ}$S to 60$^{\circ}$S, scaling the TEXES spectra to match the continuum of the CIRS spectrum.  We assume that this region of the troposphere has not changed over the seven years covered in this study, or since the Cassini observations of 2000. The obliquity of Jupiter is 3$^{\circ}$, which produces no discernible seasonal behaviour in either the troposphere or stratosphere \citep[e.g.][]{1990Icar...83..255C}. This is also outside of the range of tropospheric latitudes where belt/zone variability is observed \citep{GRL:GRL55963}. Table \ref{tscalings} lists the multiplicative scaling factors used for each observing run. These are shown in Figure \ref{scalings}. The TEXES observations in January 2017 stand out as particularly curious, as both acetylene and ethane required a scaling inconsistent with the previous observing runs. In addition, both September 2012 and January 2017 required a scaling of the methane band -- they were scaled to the mean methane radiance of the other set of observations, which appeared well calibrated. The September 2012 data used a set of non-nominal detector bias and gain settings, producing a strong non-linearity of response, whereas the January 2017 data anomaly was caused by erroneous temperature readings for the flat-field card, producing over-compensating flat-fields

\cite{2016Icar..278..128F} did preform a scaling on the 1247 cm$^{-1}$ channel, but found it made only a small difference on the retrieved temperatures. Additionally, the 1247 cm$^{-1}$ band lacked continuum emission clear from overlying methane emissions so we were unable to produce a scaling factor for this band, and it is assumed to be 1.0, apart from at the dates mentioned.

Figure \ref{scalings} shows that the calibration corrections are constant with time, apart from January 2017, indicating that this is a systematic issue in the TEXES calibration pipeline, related to the non-linear effect outlined above. 


We also adopt a shift-and-stretch fit of the wavelength scale (i.e. a linear fit: position and scale) to better match the rest wavelength of the hydrocarbon emissions, subtracting the Doppler shift of the planet, in the same manner as \cite{2014Icar..238..170F, 2016Icar..278..128F}. Once the spectra are transformed into the rest frame of the hydrocarbon emissions, they are interpolated so that each observing run is on the same wavenumber scale as was observed on December 2014.



Here, we used the telluric absorption as a measure of the quality of the data, directly linking it to the uncertainty. When the telluric absorption is high, the signal-to-noise becomes low, which produces a large uncertainty. The telluric absorption is shown as purple in Figure \ref{all_spectra}. We estimate the error on the observed intensity,  $\Delta I$, by the following expression: 
\begin{equation}
\Delta I = 0.08 \times \frac{ I_{max}}{ S_n}
\end{equation}
where $I_{max}$ is the peak intensity in a particular wavelength band and $S_n$ is the fractional telluric transmission, defined as unity when the Earth's atmosphere lets through all of the light emanating from Jupiter, and zero when all of the light is absorbed. The 0.08 multiplier was determined by measuring the noise floor away from the disk of Jupiter. 


Figure \ref{all_spectra} shows the mean equatorial spectrum for each wavelength band and each epoch, with the scaling factors of Table \ref{tscalings} applied. The black lines are the observed intensities, and the grey shaded regions indicate the uncertainty in the emission. The purple lines indicate the telluric absorption, which ranges from significant in the 744 cm$^{-1}$ band to almost nonexistent in the 819 cm$^{-1}$ band. When the telluric transmission is small, i.e. approaching 0\%, the the uncertainty in the observed radiances becomes large, as outlined above. 

For our retrievals we picked a wavelength range for each band that was relatively un-obstructed by telluric absorption (purple lines in Figure \ref{all_spectra}) for all of the observing runs. For acetylene we used the wavelength range 743.1 to 743.5 cm$^{-1}$, for ethane we used 816.5 to 820.0 cm$^{-1}$, and for methane we used 1245.0 to 1247.5 cm$^{-1}$.  These wavelength regions are indicated by vertical dashed lines in Figure \ref{all_spectra}. For each set of observations, we limit the latitude range to $\pm$70$^{\circ}$, avoiding most of the polar aurora, regions with chemistry driven by impact ionisation. 

\subsection{Maps of stratospheric hydrocarbons}

Figure \ref{maps} shows longitude--latitude maps of acetylene, ethane and methane for each of the usable observing dates listed in Table \ref{observations}, between September 2012 and January 2017, with the scalings of Table \ref{tscalings} applied. The acetylene and ethane maps have the tropospheric background subtracted, by subtracting the continuum upon which the hydrocarbon features sit, and then summing the discrete line features. In addition, in order better to show the observed brightness across all latitudes, the ethane map has been flattened by subtracting off a cosine curve that approximately follows the general trend in the emission. This has the effect of removing the large-scale latitudinal distribution and highlighting smaller changes over time. 

There is a brightening in all three wavelength bands at about about 30$^{\circ}$N, appearing in December 2014. This is above the tropospheric North Temperate Belt, seen between $\sim$24$^{\circ}$ and $\sim$31$^{\circ}$ latitude \cite[NTB,][]{2016Icar..278..128F}. The enhancement subsequently dissipates over the following few years, but still evident in January 2017, as previously noted by \cite{GRL:GRL55801}.   

The maps in Figure \ref{maps} have been scaled to show the mid-to-low latitude emissions, rendering the auroral emissions about the poles saturated. Still, they show a remarkable degree of temporal variability. This set of observations begin near northern summer solstice at Jupiter (March 2012), and end near southern summer solstice (May 2018). This change in viewing geometry is likely to contribute to the observed changes. Note that the correct emission angles were used in the retrievals, outlined in the following section. The analysis of the auroral emissions as observed by TEXES is outside the scope of this study.

\begin{figure}[h]
\centering
\includegraphics[width=1.2\textwidth]{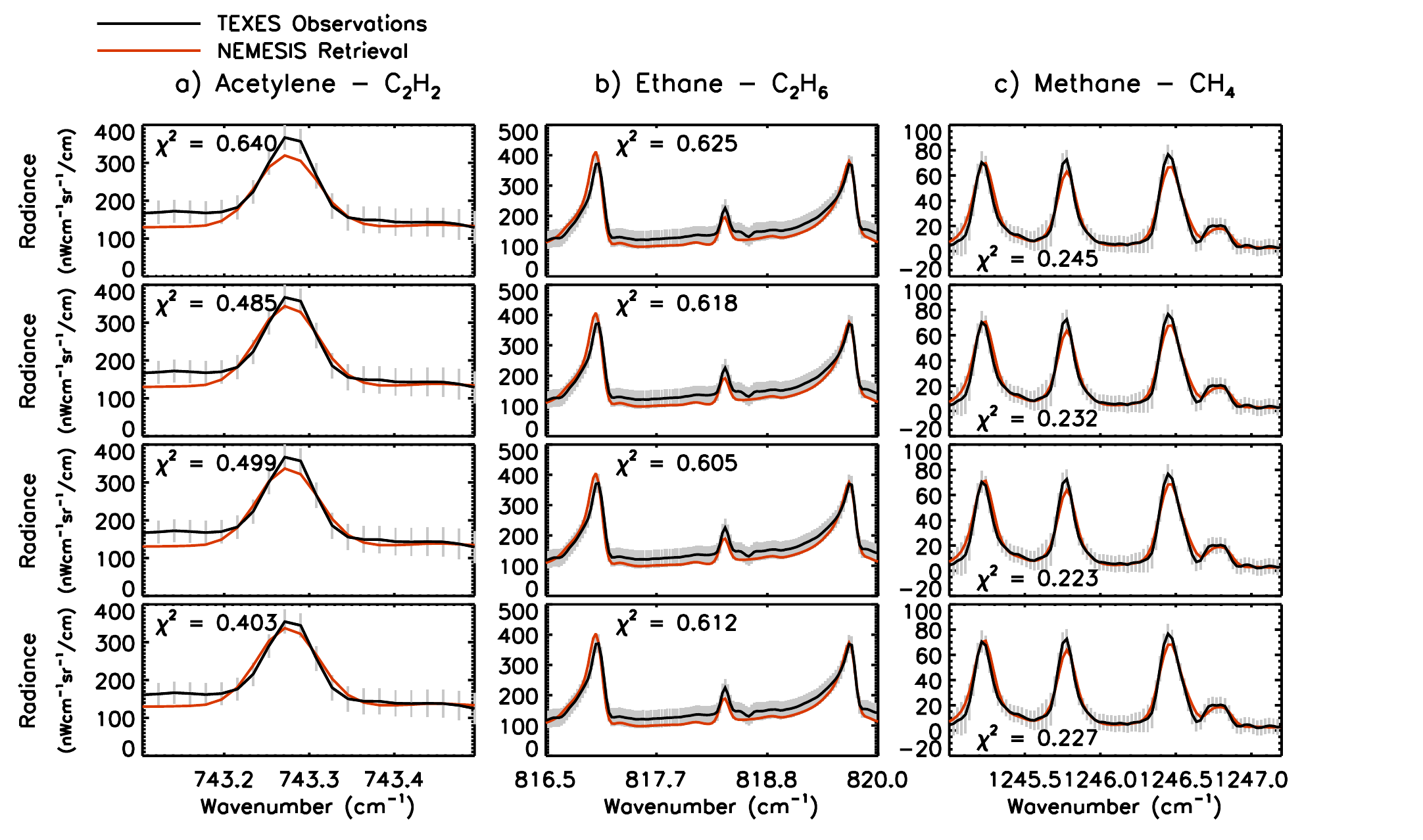}
\caption{The spectra of acetylene, ethane, and methane from December 2014, at latitude 25$^{\circ}$N for each of the four hypotheses. The first row is hypothesis A (uniform distribution), second row is hypothesis B (symmetric distribution), third row is hypothesis C (asymmetric distribution), and the bottom row is hypothesis D (distributions change with latitude and time). The observations have been scaled by the factors listed in Table \ref{tscalings}. \label{ABCDfits} }
\end{figure}

\subsection{NEMESIS}

We invert the observed TEXES spectra to determine vertical profiles of temperature and abundances, in order to investigate how these change over time. To do this we use the NEMESIS radiative-transfer and retrieval code \citep{2008JQSRT.109.1136I}. Inherent in the optimal-estimation process is an {\it a priori} knowledge of the atmospheric system in the form of vertical profiles of temperature and composition based on previous observations or model predictions. The NEMESIS retrieval produces a best fit to the observed spectra. In particular, the a priori uncertainty on the temperatures ensures that the final retrieved profile is smooth and realistic. The strength of mid-infrared emission from acetylene and ethane depend both on their abundance and the temperature of the line-forming region, whereas methane is vertically well-mixed below its homopause, and so we can use methane emission as a sensor of stratospheric temperature. 

The temperature and abundance measurements remain degenerate. The retrievals can thus be done in either two stages, first temperature, then abundances, or in a single stage. If we do the former we would rely on the fact that the retrieved temperatures from the methane band were an accurate representation of the temperatures in the ethane and acetylene line forming regions, which may not be the case. By retrieving the temperature and abundance simultaneously the degenerate effects are properly accounted for in the resulting uncertainties on the retrieved profiles. Using NEMESIS we disentangle the contribution of temperature, acetylene and ethane abundances.


We use the same a priori atmosphere as \cite{2016Icar..278..128F}, which is derived from Cassini CIRS observations \citep{2007Icar..188...47N, 2009Icar..202..543F}, in addition to the same set of spectral data and k-tables. For each NEMESIS retrieval, the temperature was allowed to stray a maximum of 5 K from the a priori temperature profile. This corresponds approximately to the range of variability observed at the 1 mbar level in the stratosphere of Jupiter \citep{2010P&SS...58.1667N}.


\cite{2016Icar..278..128F} calculated the contribution functions for each of the medium resolution TEXES channels used here. The methane band at 1247 cm$^{-1}$ samples pressures between 1 and 10 mbar, the acetylene band at 744 cm$^{-1}$ samples between 5 and 100 mbar, and the ethane band at 819 cm$^{-1}$ samples the stratosphere between 0.5 and 50 mbar. While our observations could contain sufficient information to determine the vertical structure of acetylene and ethane, we have chosen to fix the vertical structures to the a priori, and to simply scale them in the retrievals. 

At the poles, chemistry driven by auroral particle precipitation, centred about the offset magnetic field, generates elevated temperatures and modified hydrocarbon distributions that are highly localised in latitude and longitude \citep[e.g.][]{GRL:GRL56012}. In order to exclude emission associated with the auroral process, we limit the latitude range of our retrievals to be $\pm$70$^{\circ}$. 

The observed TEXES acetylene, ethane, and methane spectra were used to retrieve zonally-averaged atmospheric profiles of acetylene, ethane and temperature to examine if and how these parameters change over time. Spectral fits of acetylene, ethane and methane retrieved by NEMESIS from the December 2014 dataset are shown in Figure \ref{ABCDfits}.

%

\begin{figure}[h]
\centering
\includegraphics[width=1.2\textwidth]{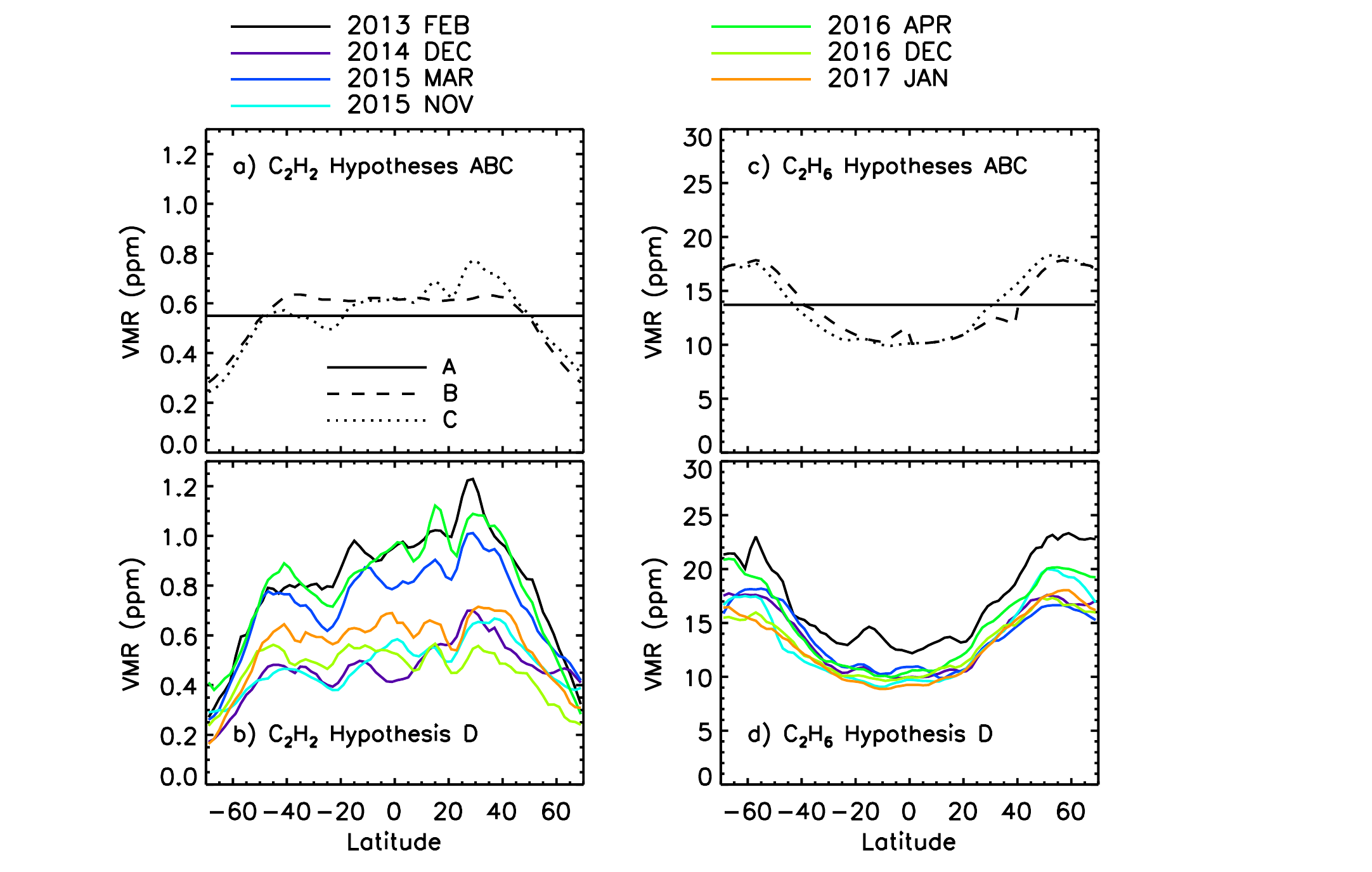}
\caption{The acetylene and ethane volume mixing ratio (VMR) at 1 mbar for each of the hypothesis tested in this study. For hypothesis A, B, and C, only the temperature is retrieved, keeping the hydrocarbon distributions fixed, where the solid line is hypothesis A, the dotted line is hypothesis B, and the dashed line is hypothesis C. For hypothesis D, temperature is retrieved along with acetylene and ethane.\label{ABCD}}
\end{figure}

\section{Analysis \label{secanalysis}}

Previous examinations of Jupiter's stratospheric emission have typically allowed temperatures and chemical abundances to vary freely to produce good fits to the data \cite[e.g.][]{2007Icar..188...47N, 2010P&SS...58.1667N}. Here, the TEXES data are subject to calibration changes from run to run, weather changes on Earth, degeneracies in the retrieval of temperature and composition, and the issue that the abundance measurements are extremely sensitive to any systematic or random uncertainties propagating through the temperature measurements. Therefore, we simply don't take the `full' retrieval at face value, and instead we develop four distinct hypotheses. One of these is the full retrieval, whilst the others represent simplified  scenarios for the distribution of the hydrocarbons. 

We will investigate whether the time-series of NASA IRTF TEXES observations can distinguish between four distinct hypotheses that describe the zonally-averaged distribution of both acetylene and ethane over time:

\begin{itemize}
\item {\bf Hypothesis A:} The abundance does not change with latitude or time, and is therefore latitudinally uniform for all dates.  If this were the case, then the variable emission seen in Figure \ref{maps} would be generated by temperature changes alone.

\item {\bf Hypothesis B:}  The abundance is {\it symmetric} about the jovigraphic equator and does not change with time. 

\item {\bf Hypothesis C:} The abundance is {\it asymmetric} about the jovigraphic equator and does not change with time. 

\item {\bf Hypothesis D:}  The abundance changes with both latitude and time. 

\end{itemize}

The zonally-averaged meridional profiles of the volume mixing ratio (VMR) for acetylene and ethane at 1 mbar for each of these hypotheses can be seen in Figure \ref{ABCD}. The acetylene zonal profile for hypothesis A (uniform) is the average value from all epochs and all latitudes of the profiles of hypothesis D (time-variable). For hypothesis B (symmetric), the mean profile of Figure \ref{ABCD}b was added to the mirrored mean profile of Figure \ref{ABCD}b, then averaged and smoothed between 40$^{\circ}$N and 40$^{\circ}$S. The profile for hypothesis C (asymmetric) is the mean of the profiles of hypothesis D (time-variable). For hypothesis D (time-variable) we retrieve temperature, acetylene, and ethane for all years, whilst for hypothesis A (uniform), B (symmetric), and C (asymmetric) we only retrieve the temperature, keeping the acetylene and ethane distributions fixed according to Figure \ref{ABCD}a and Figure \ref{ABCD}c. The same process was applied to derive the hypotheses distributions for ethane -- see Figure \ref{ABCD}c and \ref{ABCD}d.

\begin{figure}
\centering
\includegraphics[width=0.9\textwidth]{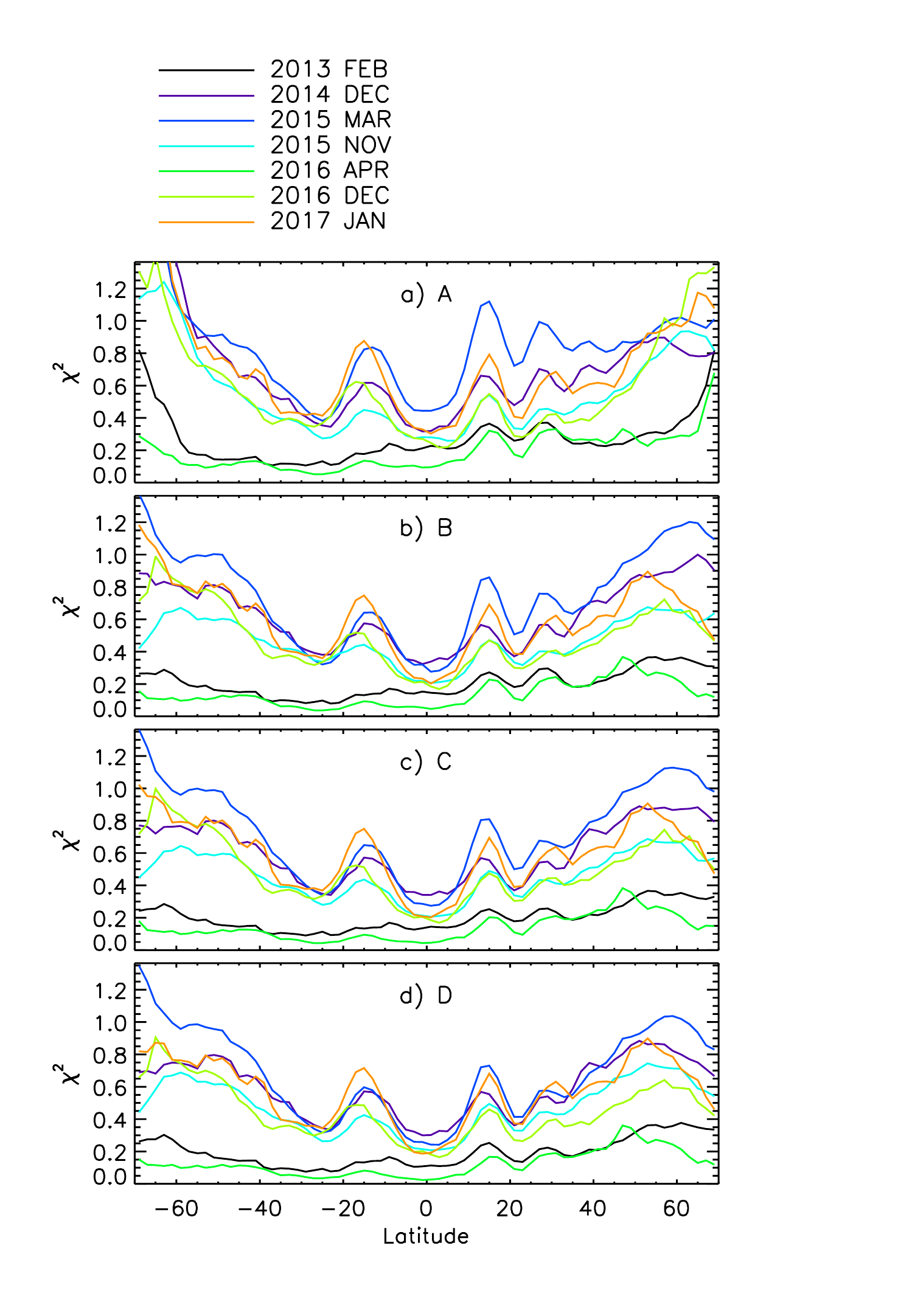}
\caption{The normalised goodness of fit parameter, $\chi^2$ as a function of latitude for the acetylene channel (744 cm$^{-1}$) from the NEMESIS retrievals of the NASA IRTF TEXES data. The results for hypotheses A, B, C and D are respectively shown in the 1st, 2nd, 3rd and 4th panels. The value of the $\chi^2$ is largely driven by signal-to-noise, and a direct comparison of the absolute $\chi^2$ is meaningless. \label{ABCDchi}}
\end{figure}

The dotted line in Figure \ref{ABCD}a shows the average of the acetylene profiles in Figure \ref{ABCD}b, with a peak volume mixing ratio of about 0.8$\pm$0.2 ppm at 30$^{\circ}$N, and a value of about $\sim$0.25$\pm$0.05 ppm at $\pm$70$^{\circ}$. Similarly, the dotted line in Figure \ref{ABCD}c shows the average of the ethane profiles in Figure \ref{ABCD}d, showing a minimum VMR of $\sim$10$\pm$0.6 ppm at the equator, and $\sim$18$\pm$1.1 ppm at $\pm$50$^\circ$.

The NEMESIS retrieval algorithm calculates fits to the observed spectra, from which we can calculate the normalised goodness of fit, $\chi^2$, defined as:
\begin{equation}
\label{eqcomp}
\chi^2 = \frac{1}{N} \sum_{i = \lambda_{start}}^{\lambda_{end}} \frac{(fit_i - data_i)^2}{data_i}
\end{equation}
where $N$ is the number of spectral bins, $\lambda_{start}$ and $\lambda_{end}$ are the wavenumber bounds of interest, $fit_i$ and $data_i$ is the fitted and observed radiance in wavelength bin $i$ respectively. Since we are interested in what the differences in $\chi^2$ are between our hypotheses in both the acetylene and ethane spectra, individual $\chi^2$ are calculated for the 744 cm$^{-1}$ and 819 cm$^{-1}$ channels. The $\chi^2$ parameter is strongly dependent on the signal-to-noise of the observed spectrum, and larger $\chi^2$ means a larger discrepancy between the observed spectrum and the retrieved model spectrum. In turn, the signal-to-noise is dependent on a number of parameters, e.g. observing conditions, dark currents, and readout noise. It is therefore difficult to compare spectral fits from different observing runs which have different signal-to-noise. This is illustrated in Figure \ref{ABCDchi}, which shows the retrieved $\chi^2$ for each hypothesis and observing run for the acetylene channel. The March 2015 has almost always the highest $\chi^2$, meaning it was more challenging to fit with our spectral model. In contrast, the April 2016 run has the lowest $\chi^2$, indicating a good signal-to-noise. All of the $\chi^2$ profiles in Figure \ref{ABCDchi} show an increase at about $\pm$15$^{\circ}$, and towards the poles -- this is also driven by signal-to-noise, as a noisier spectrum produces a larger $fit-data$ residuals. These regions also appear much darker in the maps shown in Figure \ref{maps}, indicating a lower signal (brightness), and thus mismatches between observed and modelled spectra form a larger proportion of the observed spectra.  

The spectra shown in Figure \ref{ABCDfits} highlight how the quality of fit changes according to which hypothesis is used. We show retrieval models for acetylene and ethane from December 2014, at a latitude of 30$^\circ$N, for each of the four hypothesis. For acetylene, we see a general decrease in the $\chi^2$ as we move from hypothesis A to D, indicating a better fit to our observed spectrum.  

\begin{figure}
\centering
\includegraphics[width=1.1\textwidth]{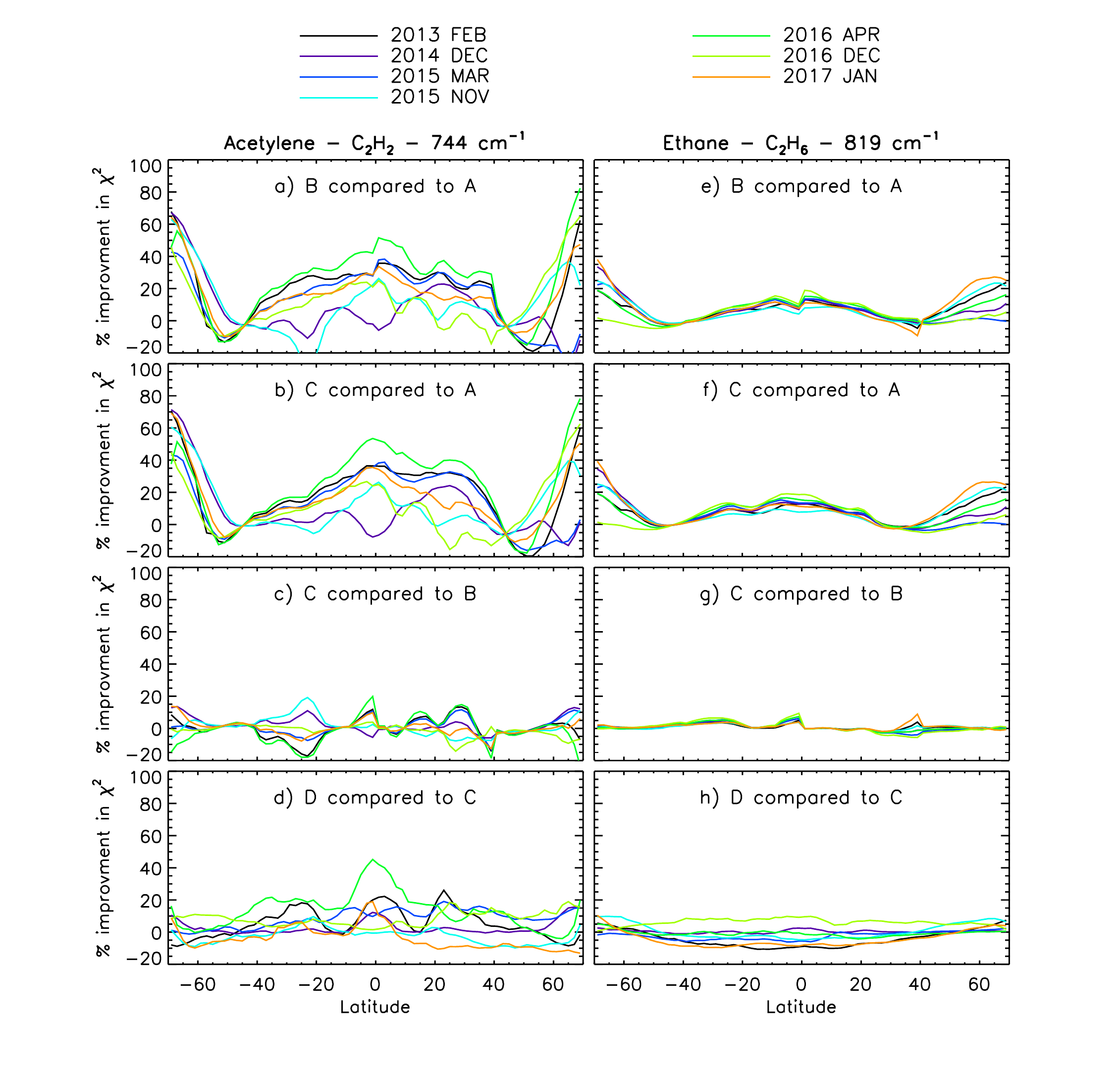}
\caption{The percentage change in $\chi^2$ between the different NEMESIS retrivals for both the acetylene and ethane wavelength channel. We compare the quality of fit for each of the four different hypotheses shown in Figure \ref{ABCD}. \label{ABCDtest}}
\end{figure}

In order to compare retrieved $\chi^{2}$ of different observing runs, we define the change in the quality of the fit as a percentage difference of two sets of retrieved $\chi^2$, $\Upsilon$:
\begin{equation}
\Upsilon_{mn} = \frac{\chi^2_m - \chi^2_n}{\chi^2_m} \times 100 \%
\end{equation}
where $\chi_m^2$ and $\chi^2_n$ represents the $\chi^2$ from NEMESIS retrievals for hypothesis $m$ and $n$. Figure \ref{ABCDtest} shows the percentage improvement in $\chi^2$, comparing the quality of the fit for each of the hypotheses above. The different panels in Figure \ref{ABCDtest} will be discussed in the following section.

\begin{figure}
\centering
\includegraphics[width=1.2\textwidth]{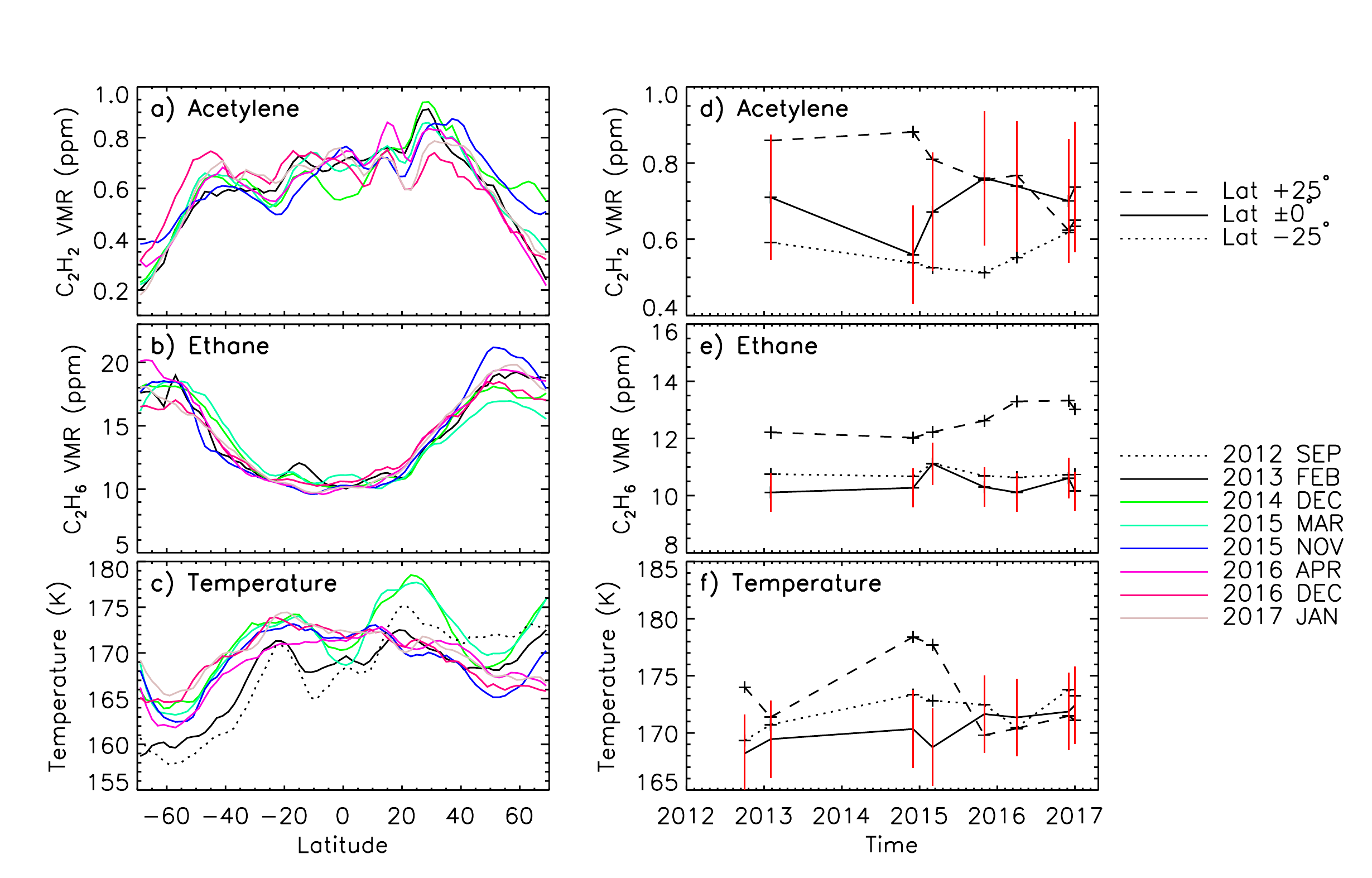}
\caption{The collapsed acetylene and ethane abundance profile at 1 mbar, together with the zonal-mean temperature profiles at the same pressure. For each observing run, the distribution was rescaled to have the same average abundance equatorward of $\pm$60$^{\circ}$ latitude. In the right panels, for ease of viewing, error bars are only given for the equatorial data, but are of equal amplitude at all latitudes. Note that for the September 2012 data insufficient  acetylene or ethane data are available, and so only the temperature retrieved from methane is shown.  \label{ABCDadj}}
\end{figure}

\section{Results \label{secresults}}

\subsection{Temperature \label{sectemp}}

Before discussing the distributions of the hydrocarbons, we begin by reviewing the  temperatures retrieved for each observing run. Figure \ref{ABCDadj}c shows the `raw' temperature retrievals versus latitude at 1 mbar, revealing variability at all latitudes of about $\pm$5 K with an uncertainty of about $\pm$3 K, with Figure \ref{ABCDadj}f showing the time-series. Note that the temperature from the September 2012 observations was retrieved using only the methane band at 1247 cm$^{-1}$, rather than all three spectral channels used in subsequent retrievals. There is a clear heating event in December-2014 and March-2015 data at 25$^\circ$N. This heat has mostly dissipated by November 2015, although remnants of this meteorological event are still visible during later observations -- see Figure \ref{maps}. This heating at northern mid-latitudes may coincide with the expansion of the North Equatorial Belt \citep[NEB,][]{GRL:GRL55801}, confirming the elevated 1 mbar temperatures identified by \cite{2016Icar..278..128F} in the December 2014 TEXES data. The consequences of this mid-latitude heating event on the hydrocarbon distributions will be explored in the following sections. In addition, the low latitudes show some temperature variability associated with Jupiter's Quasi-Quadrennial oscillation -- this phenomenon, and the impacts on the vertical hydrocarbon distributions, will be discussed in Section \ref{secqqo}.  


\subsection{Acetylene -- C$_2$H$_2$ \label{secacetylene}}

Figure \ref{ABCDtest}a shows the percentage improvement in the goodness of fit, $\Upsilon_{AB}$, for the acetylene channel at 744 cm$^{-1}$, comparing a latitudinally uniform distribution (A) to one that is symmetric about the equator (B). There is a positive change in $\Upsilon_{AB}$ providing strong evidence that hypothesis B better describes the observations than hypothesis A: namely, that we must allow the abundances of acetylene to vary with latitude in order to fit the data, and varying temperature alone is insufficient.

At mid-latitudes in Figure \ref{ABCDtest}, both in the northern and southern hemisphere, the $\Upsilon_{AB}$ parameter tends towards zero. The reason for this is shown in Figure \ref{ABCD}a, showing the abundance of acetylene and ethane as a function of latitude for each of the hypotheses: the volume mixing ratio of hypothesis A and B, are the same at a latitude of $\pm$50$^{\circ}$. 


Figure \ref{ABCDtest}b shows the $\Upsilon_{AC}$ comparison between a flat (A) and asymmetric (C) acetylene distribution. Similarly as above, the asymmetric distribution is a much better fit to our observations. Hence, we can rule out a distribution of acetylene that is constant with latitude.  

Figure \ref{ABCDtest}c shows the comparison between the symmetric (B) and asymmetric (C) hypothesis. Here, a deviation from $\Upsilon_{BC} = 0$ indicates that the asymmetric model at some latitudes and epochs improves and at others worsens the quality of the fit with respect to a symmetric model. The deviation from zero occurs at latitudes where the two abundance profiles are different (see Figure \ref{ABCD}a). This indicates that at some latitudes, the distribution of acetylene cannot be symmetric about the equator, and that the asymmetric hypothesis is a better description of the data than the symmetric model.

We now compare the asymmetric model to the one where acetylene and ethane are allowed to vary with latitude and with time, $\Upsilon_{DC}$. The NEMESIS runs associated with hypothesis D retrieves temperature {\it and} scaled abundances of ethane and acetylene, with the retrieved family of acetylene latitude profiles being shown in Figure \ref{ABCD}b. Figure \ref{ABCDtest}d shows, unsurprisingly, that all epochs benefit to some degree by having more degrees of freedom in the retrievals. However, as seen in Figure \ref{ABCD}b there remains a systematic offset in the VMR of the retrieved acetylene abundance of each observing run. Given that these offsets are systematic, they are likely related to changes in the radiometric calibration from date to date, and the general spread may be a better representation of our uncertainty in the absolute abundances being retrieved, than the ones furnished by NEMESIS. We assume that the acetylene and ethane  abundances do not increase or decrease systematically at all latitudes over these short timescales. Therefore, what could be interpreted as evidence for temporal changes in Figure \ref{ABCDtest}d may be just be a result of these offsets. An attempt to rectify this is performed in Section \ref{sectempvar}.

\subsection{Ethane -- C$_2$H$_6$} 

Figure \ref{ABCDtest}e shows $\Upsilon_{BA}$, the comparison between the uniform distribution (A) and symmetric distribution (B) of ethane, with Figure \ref{ABCDtest}f showing the comparison between the flat (A) and the asymmetric model (C), $\Upsilon_{CA}$. The positive improvements at all latitudes, apart from at the ones where the three hypotheses have the same abundance, is evidence that we can reject the hypothesis of an ethane abundance that is constant with latitude. 

Next, we compare the symmetric (B) to the asymmetric (C) hypotheses, $\Upsilon_{BC}$, shown in Figure \ref{ABCDtest}g. Here, there are very small  improvements to the quality of the fit, particularly at the equator and at $\sim$30$^{\circ}$N, which is where the two ethane abundance profiles in Figure \ref{ABCD}c are different. However, these changes are small, and we are unable to state unequivocally that an asymmetric hypothesis performs better than an symmetric one. Therefore, from the TEXES data-set used here, a distribution of ethane symmetric about the equator and invariant with time appears to be a good representation of the data. 


Lastly, we compare the asymmetric hypothesis (C) to the hypothesis where ethane and acetylene are allowed to vary (D), $\Upsilon_{CD}$, shown in Figure \ref{ABCDtest}h. These profiles show slightly larger departures from $\Upsilon = 0$ than  $\Upsilon_{CB}$ in Figure \ref{ABCDtest}h by virtue of having more degrees of freedom, but they are essentially flat, and we see no evidence that ethane varies over time.   

In summary, this work shows that the TEXES dataset favours an asymmetric and changing distribution of acetylene, but a symmetric and unchanging distribution of ethane.

\subsection{Temporal variability \label{sectempvar}}

The retrieved abundances of acetylene and ethane in Figure \ref{ABCD}b and \ref{ABCD}d shows that individual retrievals for the same gases all share a common distribution but have systematic offsets. For example, the acetylene abundance in December 2014 and March 2015 have a very similar shape, whilst being offset approximately 0.3 ppm across all low- to mid-latitudes. Such a systematic offset across the entire planet is unlikely to be physical, and masks smaller-scale local variations in the gaseous composition (e.g., related to wind, wave and atmospheric circulation).  Such systematic biases are more likely due to small errors in the absolute calibration in all of the TEXES channels, or that our method of scaling the TEXES observations to Cassini CIRS is not entirely sufficient. In order to assess how these abundances change over time we scale the abundance profiles to compensate for these calibration offsets. We make the assumption that the spatially averaged VMR equatorward of $\pm$60$^{\circ}$ latitude does not change significantly over time, and then scale each observing run to have the same spatially averaged VMR. The scaled VMR of these species at 1 mbar can be seen in Figure \ref{ABCDadj}a and \ref{ABCDadj}b plotted versus latitude. Figure \ref{ABCDadj}c shows the retrieved temperature at the same pressure level and demonstrates some differences of the order of 5 K from date to date.  Whether this is real, or related to differences in the observing conditions, is hard to say. The right column shows the temporal evolution of acetylene (\ref{ABCDadj}d), ethane (\ref{ABCDadj}e) and temperature (\ref{ABCDadj}f) at three different latitudes, 25$^\circ$S, 0$^\circ$, and 25$^\circ$N. The red error bars shows the error associated with the retrievals at the equator -- the other latitudes have errors of similar magnitude.





The collapsed acetylene distribution in Figure \ref{ABCDadj}a shows that there is variability in the retrieved distributions with time, although error bars for these retrievals are on the order of $\pm$0.15 ppm. The time-series in Figure \ref{ABCDadj}d shows a distribution that starts off clearly asymmetric, and 2013-14, with northern latitudes demonstrating higher acetylene abundances than southern, but in moving towards the last set of observations in January 2017 the distribution is clearly becoming more symmetric. The error bars are substantial but the deviations over time are not random, instead appearing at $+25^\circ$ to fall over the 6-year period of the dataset. This changing acetylene asymmetry is consistent with our conclusions in Section \ref{secacetylene}.



The collapsed ethane distribution in Figure \ref{ABCDadj}b shows less variability than the acetylene, with an uncertainty of about $\pm$1 ppm. The time-series in figure \ref{ABCDadj}e shows very little variability, apart from a suggestive increase in the equatorial abundance between November 2015 and January 2017, which remains inside the levels of uncertainty. 


\subsection{Quasi-Quadrennial Oscillation \label{secqqo}}

Our data-set covers about half a jovian year, slightly longer than the 4-year cycle of the QQO, and so we can examine to what extent the medium-resolution TEXES data can address this cyclical behaviour in the stratosphere. \cite{cosentino17} used higher-spectral resolution TEXES observations to show that an alternating pattern of warm and cool thermal anomalies descends downwards over the equator over time, with an amplitude of about $\sim$5 K.   



The 1 mbar temperature time-series shown in Figure \ref{ABCDadj}f does not readily reveal the QQO cycle at the equator (solid line). This is because the temperature differences seen at 0$^{\circ}$ latitude (Figure \ref{ABCDadj}c) are mainly driven by the small calibration offsets present in the TEXES medium-resolution dataset, generating a temperature scatter that has a similar magnitude to the expected QQO temperature amplitude. In an attempt to correct for this, we calculate the temperature anomaly, as outlined below. 

\begin{figure}
\centering
\includegraphics[height=40pc]{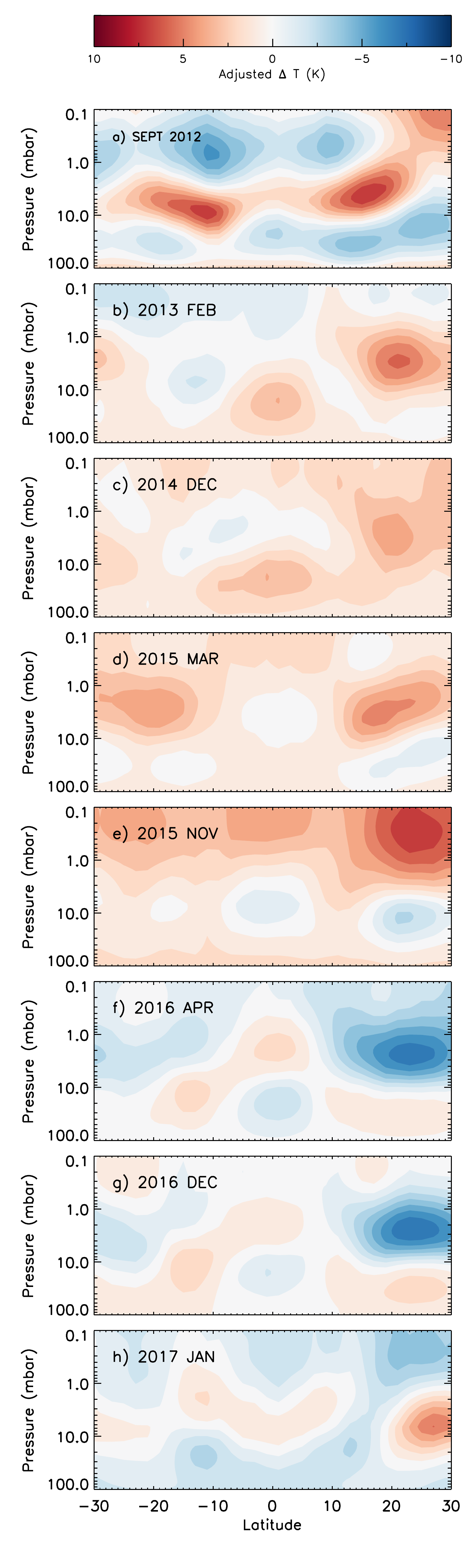}
\caption{The temperature anomalies for each epoch as a function of latitude and pressure. These distributions are the difference from the mean temperature for all epochs. We have normalised the temperature maps to account for calibration offsets. The errors on the retrieved temperatures is $\sim$3 K.  The Quasi-Quadrennial Oscillation temperature anomaly can be seen moving downwards at the equator at any particlar pressure over time, alternating between positive and negative $\Delta T$.\label{QQOadj} }
\end{figure}

The temperature anomaly is calculated by first rescaling the temperature to a common average, using only the temperatures contained within 10$^{\circ}$S to 10$^{\circ}$N. The anomaly is then the difference between the rescaled temperature for a particular observing run and the average rescaled temperature. This procedure attempts to correct for the calibration offsets, in a similar manner as for acetylene and ethane in Figure \ref{ABCDadj}a and \ref{ABCDadj}b. Figure \ref{QQOadj} shows the the stratospheric temperature anomalies over time, as a function of pressure and latitude. This method shows that we still see similar temperature fluctuations as \cite{cosentino17}, even though the spectral resolution of our data is a factor of 5-8 lower. Since we observe the QQO temperature oscillation to be present in the medium-resolution TEXES data analysed here, we can now address to what extent the QQO affects the abundances of acetylene and ethane. For example, in a zonal time-series, if a minimum in temperature coincides with a minimum abundance of acetylene, that would suggest an upwelling of cold acetylene-depleted air, a secondary circulation pattern associated with the QQO cycle. 

The collapsed acetylene and ethane abundances in Figure \ref{ABCDadj}d, \ref{ABCDadj}e  are corrected for the TEXES calibration offsets (see Section \ref{sectempvar}), and the time-series in Figure \ref{ABCDadj}d and \ref{ABCDadj}e shows how these species vary over time. The acetylene distribution in Figure \ref{ABCDadj} does show a minima at the equator in late 2014/early 2015, whilst ethane appears unchanging over time. This minimum in the 1 mbar acetylene abundance may be connected to the temperature minimum in the QQO temperature anomaly, consistent with an upwelling of cold acetylene deficient air. However, the variability is completely contained within the retrieval errors and so this dataset provides no concrete evidence that the QQO cycle affects the acetylene and ethane abundances. This analysis is at the limit of what we can do with these data, and in order to investigate to what extent the QQO cycle modulates the hydrocarbon distributions we need to employ higher spectral resolution observations. 

\section{Discussion, \label{secdiscussion}}

\subsection{Distributions of acetylene and ethane}

In analysing the mid-infrared TEXES observations obtained between September 2012 and January 2017 we have confirmed that the two hydrocarbons are decoupled from one another, with acetylene decreasing towards the pole, and ethane increasing toward the pole. This decoupling is a permanent feature of Jupiter's stratosphere, persisting throughout the 6-year observing period and was present also during Cassini's flyby of Jupiter in 2000 \citep{2007Icar..188...47N}. Additionally, we have found that the distribution of acetylene is asymmetric about the jovigraphic equator, being more abundant at mid-northern latitudes than at mid-southern latitudes. The acetylene distribution moves from being asymmetric in 2012 to being near-symmetric in 2017, although the error bars accrued from the retrieval and subsequent scalings are on the same order as the observed asymmetry. 


The fact that the acetylene and ethane distributions are so different is perplexing. The abundances of these hydrocarbons are governed predominantly by the solar flux, which has the highest energy density about the equator. Additional more complex sources and sinks, such as ion-neutral chemistry, subsequently alters these distributions. The hydrocarbons are  transported by two principal means: 1) {\it Diffusive processes}, where molecular and turbulent diffusion acts in the vertical direction, whilst meridional turbulent diffusion act in the latitudinal direction \citep[e.g.][]{2002Icar..159..112L} 2) {\it Advective processes}, such as the movement of a global circulation cell. If acetylene and ethane had very short lifetimes, say hundreds of days, we may expect a poleward decrease for both species, following the daily averaged meridional solar flux variation \citep{2004jpsm.book..129M, 2005ApJ...635L.177L, 2013P&SS...88....3Z}. If instead they had very long lifetimes, say hundreds of years, then the global circulation of upwelling at the equator and downwelling at the pole \citep{1990Icar...83..255C, 2005ApJ...635L.177L}, would form a distribution increasing towards the pole. The peak production of ethane occurs at $\sim$1 $\mu$bar \citep{2005JGRE..110.8001M}, and it is transported to the lower stratosphere via diffusion. The fact that ethane increases towards the pole despite having a peak production rate at the equator indicates that the depletion rates are slower than the rates at which ethane is transported towards the pole. Observations of Jupiter's stratosphere following the Shoemaker--Levy 9 impact \citep[e.g.][]{1999Icar..138..141F, 2002Icar..159..112L, 2004Icar..170...58G} could be explained with only meridional diffusion, although \cite{2006Icar..184..478L} required both meridional diffusion and advection to explain the abundance distributions of HCN and CO$_2$. At Saturn, turbulent diffusion has been shown to be insufficient to explain the increase of ethane abundance with latitude \citep{2007plat.work...85M}. Hydrocarbons in Jupiter's stratosphere are likely mixed by both advective and diffusive processes, and future modelling work will be required to connect the atmospheric chemistry with diffusion and general circulation in the stratosphere.


Acetylene, on the other hand, has a distribution centred about the equator. If acetylene has a lifetime much shorter than the advection time-scale, with upwelling at the equator and downwelling at the pole, then we would expect it to have high abundances close to where it is created \citep{2005ApJ...635L.177L, 2013ApJ...767..172Z}, which is what is observed. In addition, it seems likely that these distributions are stable over long time periods given that the distribution is similar to that observed by CIRS \citep{2007Icar..188...47N}. Whilst \cite{2005ApJ...635L.177L} suggested that the lifetime argument in isolation is likely too simplistic, the lifetimes calculated by \cite{2010P&SS...58.1667N} are hundreds of days for acetylene, and hundreds of years for ethane. If the time-scale for advecting these species globally is intermediate between the acetylene and ethane chemical timescales, say a few years, then the long-lived ethane could be showing the abundance redistributed by global scale advection, meridionally transporting material from the equator, where the hydrocarbon are produced, to the poles. The short-lived acetylene would not survive these long dynamical cycles, and thus remaining predominantly about the equator.

It must also be noted that the poles of Jupiter are a rich source of methane chemical products, due to the ion-neutral reactions driven by particle precipitation from the magnetosphere \citep{2017Icar..292..182S}. This source of both acetylene and ethane could be evident at latitudes greater than $\pm$70$^{\circ}$, but it remains unclear whether the hydrocarbons produced at these high latitudes can be advected equatorward to produce the distributions we observe.


\subsection{Temporal evolution of acetylene}

Since the axial tilt of Jupiter is small ($\sim$3$^{\circ}$) we do not expect any extreme seasonal behaviour in the planet's stratosphere \citep{1990Icar...83..255C}. However, \cite{2010P&SS...58.1667N} observed broad-scale abundance changes in both acetylene and ethane between 1979 and 2000. Here, we are seeing small changes over a period of 5 years, between 2013 and 2017. In particular, there is a change in the symmetry of mid-latitude acetylene, beginning in November 2015, with the abundance at $+25^\circ$ decreasing and the abundance at $-25^\circ$ increasing. This suggests that the distribution is becoming more symmetric as we head towards January 2017. This can also be seen in Figure \ref{ABCDtest}c, where the December 2016 and January 2017 observing runs produce lines closest to $\Upsilon_{CB} = 0$, indicating that the asymmetric hypothesis offers no improvement on the symmetric. This evolution of the acetylene, from asymmetric to symmetric, starts when we see the temperature enhancement at 25$^{\circ}$N in December 2014 and March 2015 (Figure \ref{ABCDadj}f). The enhancement in mid-northern latitude temperatures in 2014-15 was first observed in \cite{2016Icar..278..128F}. It appears to be related to the significant longitudinal variability of the NEB seen in Figure \ref{maps}, suggestive of wave phenomena propagating in the stratosphere. The localisation of the highest brightness temperature in December 2014 near to a ragged northern edge of Jupiter's NEB prompted speculation that it could be related to dynamic activity in the troposphere, particularly that related to an expansion of the NEB \citep{GRL:GRL55801}. The stratospheric wave pattern was observable at higher spatial resolution from VLT-VISIR \citep{2004Msngr.117...12L} throughout early 2016. This TEXES dataset confirms that the wave tracked by \cite{GRL:GRL55801} had a lower amplitude than that seen in 2014-2015 by TEXES: i.e., Fletcher et al. were likely seeing the after-effects of the process that caused the thermal enhancement near 270$^{\circ}$W in 2014-15. The cooling of the perturbation over a $\sim$1-2 year timescale could be caused both by radiative cooling from stratospheric hydrocarbons, combined with mixing and homogenisation with longitude by the propagation of the wave itself.  

In this study we present data from 2012 to 2017. A longer time series is beyond the present scope of the analysis, but a similar northern wave was observed during Cassini \citep{2004Natur.427..132F, 2006Icar..185..416L, 2016Icar..278..128F} and at the same time acetylene was observed to be more asymmetric \citep{2007Icar..188...47N}. The acetylene asymmetry could be driven by a vertically propagating wave, advecting acetylene-enriched air from higher stratospheric levels to the 1-mbar level. In such a case, we would expect the acetylene asymmetry to be more prominent during periods of strong wave activity near the northern edge of the NEB. 


A wave-generated bulk transport towards lower altitudes (and lower temperatures) would also move ethane, suggesting that this too would have an asymmetry. Yet the observed ethane distribution is inconsistent with an asymmetric distribution. This could be a result of the fact that fact that acetylene has a stronger vertical gradient in the region \citep{2005JGRE..110.8001M} to which TEXES is sensitive. Thus, a small vertical displacement can have substantial consequences for the VMR perturbation at the 1-mbar level. Ethane has a shallower gradient than acetylene \citep{2005JGRE..110.8001M}, so it would take a far larger displacement to have any notable consequences on the ethane emission.



In summary, the spatial variability in the hydrocarbons appears to have three causes: 1) small-scale and short-lived variation caused by stratospheric mixing (and potentially wave activity); which strongly affects species with large vertical gradients in abundance; 2) large-scale upwelling at the equator and downwelling at the poles generate gradients that are determined by the balance between the chemical timescale and the global circulation timescale \citep{1990Icar...83..255C}; and 3) variable production via ion-neutral chemistry in the polar regions as described by e.g. \cite{2017Icar..292..182S}, followed by advection to lower latitudes.





\section{Summary \label{secsummary}}

We have analysed NASA IRTF TEXES observations obtained between 2012 and 2017. We made four distinct hypotheses about the distribution and temporal evolution of acetylene and ethane: three that are invariant with time, but have a flat (A), symmetric (B) and asymmetric (C) distribution about the equator, with the last hypothesis being that they vary across all latitudes with time. 

The TEXES data supports the hypothesis that acetylene is best described by an asymmetric distribution about the equator, that varies with time. Ethane, in contrast, is symmetric about the equator and does not vary with significantly with time, although there is some localised variability at mid-latitudes whose source is uncertain. 

We suggest that the change in the abundance distribution of acetylene is driven by stratospheric waves that mix acetylene-rich air to lower altitudes. The differences between the distributions of acetylene, which increases towards the equator, and ethane, which increases towards the pole, is likely linked to their different chemical lifetimes. Acetylene is only short lived (hundreds of days), so does not serve as a tracer of longer timescale dynamics. Ethane has a very long lifetime (hundreds of years), and is re-distributed by equator-to-pole transport that occur over very long timescales. 

\section{Acknowledgements}
This work was supported by the UK Science and Technology Facilities Council (STFC) Grant ST/N000749/1 for Melin. Fletcher was supported by a Royal Society Research Fellowship at the University of Leicester. Melin, Greathouse, Giles, Fletcher and Orton are Visiting Astronomers at the Infrared Telescope Facility, which is operated by the University of Hawaii under contract NNH14CK55B with the National Aeronautics and Space Administration.


\bibliography{refs}

\end{document}